\documentclass[12pt,aps,ajp,paper]{revtex4}   % style for Physical Review B and AJP are similar

\usepackage{amsmath}    % need for subequations
\usepackage{graphicx}   % for figures
\usepackage{textcomp}   % for figures

\begin{document}

\title{More reflectivity for the soil to counteract the global-warming of the Earth}
%Lines break automatically or can be forced with \\
\author{A. Tejedor}
 %%\altaffiliation[Also at ]{home.}  %  optional
 \affiliation{Faculty of Sciences. University of Zaragoza.}
 \email{atejedor@unizar.es}   %optional
%%\email(atejedor@unizar.es}   %optional

\author{A.F. Pacheco}
\affiliation{Faculty of Sciences and BIFI. University of Zaragoza.}

\date{\today}

\begin{abstract}
It is argued that a dedicated effort to increase the reflectivity of the surface of our planet by means of, for example, metallic plates would induce an increase in the global albedo which would counteract in part the present global-warming process of the Earth. This could alleviate the urgency of reducing the $CO_2$ emissions. The City of Zaragoza (Spain) is chosen to illustrate the likelihood of our arguments.
\end{abstract}

\maketitle

There is an almost total consensus among experts \cite{IPCC} that the climate of our planet is in a clear process of global warming.  The so-called greenhouse effect induced by the sustained emission of $CO_2$ and other gases seems to be the main factor responsible for this phenomenon. Given the very grave consequences of this process (such as the increase of the sea level, acidification of the oceans, etc.), this issue is likely to be one of the main problems facing mankind in the XXI century. Therefore, the international community has launched several initiatives to limit the emission of $CO_2$ into the atmosphere. Depending on the peculiarities and circumstances of each country, an abrupt accommodation to these norms is very hard. 
\\

In this short paper (using well known physics) it is recalled that a  somewhat modest effort to increase the reflectivity of the surface of our planet by means of metallic plates would induce an increase in the global albedo which would efficiently counteract in part the present global-warming process. This could alleviate the urgency of reducing the $CO_2$ emissions.
To illustrate these ideas, we first need a model that enables us to compute the planetary albedo as a combined effect of the atmospheric gas, the clouds and the surface.

\begin{figure}[h]
\begin{center}
\includegraphics[width=20pc]{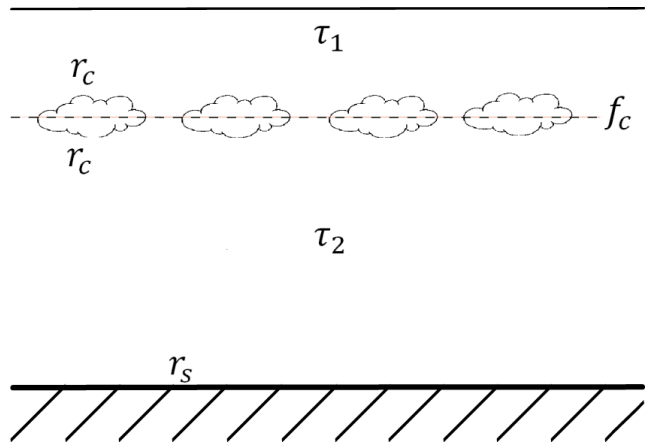}
\caption{\label{fig1} Model of short wave radiation. For details, see the text.}
\end{center}
\end{figure}

Consider the simplified model of short-wave energy balance shown in Fig.\ref{fig1}. (taken from \cite{Wallace}). In this model, we see that the atmosphere has been represented by an upper layer with transmissivity  $\tau_1$, a partial cloud layer with the fractional area $f_c$ covered by clouds, with reflectivity $r_c$ in either direction, and a lower layer with transmissivity $\tau_2$. Besides, the Earth´s surface has an average reflectivity $r_s$. It is additionally assumed that the reflectivity of the two atmospheric layers and the absorptivity of the clouds are null.

For this model, the total short wave radiation reaching the surface of the planet divided by the solar radiation incident upon the top of the atmosphere, $F_s$, is  
\begin{equation}\label{FS}
F_s= \frac{[(1-f_c)+f_c(1-r_c)]\tau_1\tau_2}{1-\tau^2_2f_cr_cr_s}
\end{equation}

and the combined albedo, $A$, is given by

\begin{equation}\label{ALB}
A= f_cr_c\tau^2_1+F_sr_s[(1-f_c)+f_c(1-r_c)]\tau_1\tau_2
\end{equation}

To fix the ideas, we will assume that\\
\\
$r_c=0.5$\\
$\tau_1=0.95$\\
$\tau_2=0.90$\\

The high values of $\tau_1$ and  $\tau_2$ correspond to the high transparency of the atmospheric gas to the visible light. The lower layer has a smaller transmissibility because of its higher molecular density. The assumption that the cloud reflectivity is 0.5 constitutes a reasonable average for the different types of clouds existing in the troposphere.

\begin{figure}[h]
\begin{center}
\noindent\includegraphics[width=20pc]{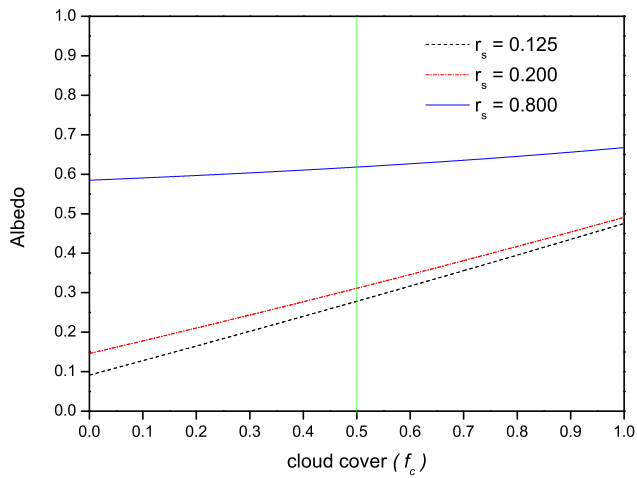}
\caption{\label{fig2} Albedo of the Earth in terms of cloud cover, $f_c$, for bare soil (dashed line), vegetation cover (dotted line) and a high-reflectivity surface (continuous line).}
\end{center}
\end{figure}

In Fig.\ref{fig2}, the albedo of the Earth resulting from the model of Fig.\ref{fig1} is represented as a function of the cloud cover $f_c$. The dashed line represents the case of  bare soil ($r_s=0.125$), the dot-dashed line corresponds to a surface covered by plants ($r_s=0.20$), and the solid line represents the case of a high-reflectivity surface ($r_s=0.80$). This reflectivity corresponds to a standard metal plate, typically of aluminium.

For an intermediate value of the cloud cover ($f_c=0.5$), $A=0.2783$ for the bare soil, and $A=0.6181$ for the metallic surface respectively. Hence the gain in the value of the Earth's albedo, when a surface of bare soil is replaced by a metallic one, is $\Delta A \approx 0.34$ .

Zaragoza in Spain is the capital city of the autonomous region of Arag\'on. Its population is 682,300, the fifth largest city in Spain.  It lies at an altitude of 199m, latitude 41\textdegree 38'1'' N and longitude 0\textdegree 53' 1'' W.

The amount of solar radiation incident at the top of the atmosphere depends on the latitude, season and time of the day. Whilst the annual mean insolation \cite{Hartmann94}, $I$, in the poles is around $150 W/m^2$ and $420 W/m^2$ at the equator, at a mean latitude like that of Zaragoza the mean insolation is around $380W/m^2$.

We will use the Giga Watt as the unit of reflected power and the $Km^2$ as the unit of surface whose reflectivity has been increased. Then, the amount of soil surface, $S$, to be covered by metallic plates to reflect to space an amount of, say, $1 GW$ is:

\begin{equation}\label{ALB}
S= \frac{1GW}{\Delta A·I}
\end{equation}

Substituting  $\Delta A= 0.34$, we obtain that at the equator, $S= 7 Km^2$, and at a mean latitude, $S= 7.7 Km^2=(2.8 Km)^2$.

The task of covering a square of side $2.8 Km$ with metallic plates does not look like an impossible task. However, now the question would be: is the elimination of 1GW solar energy by reflection an amount competitive with the scale of the heat wasted by humans? 

The {\it primary} energy consumption {\it per capita} in Arag\'on is 4.77 tons of oil equivalent per year \cite{IAEST}, this being equivalent to 205 Giga Joule per year or 6.5 $KW$.
Hence, the total power consumed in Zaragoza is of the order of 4.4 $GW$. This is the contribution of this city to the waste heat generated by mankind at the surface of the Earth. (Note that it includes both non-renewable and renewable energy resources).
Then, $1GW$ is around 23\% of the total contribution of this City. This is not so bad! 

In fact, the idea contained in this paper is not new. For example, plants reflect some of the incoming solar light back into space, and therefore replacing today´s crops with strains that reflect more sunlight could help fight global warming. This idea has been defended in \cite{Rid09}, where the temperature decrease that those crops would locally induce in the summer time has been estimated.
What we have remarked here is the importance of using a highly reflective material. In the context of the model of Fig.\ref{fig1}, the value of $r_s$ for a selected crop would be \cite{MU08} around 0.2 and as seen in Fig.\ref{fig2}, the gain in albedo would be only 0.03. Besides, our arguments are based on unquestionable energy balances which are more reliable than the temperature-change estimations.
\\

Several concluding remarks can be made. First, the strategy defended here obviously works better in places with a higher insolation, i.e. low latitudes. 
Second, so far we have taken for granted that the metallic plates would lie horizontally on the surface. If the plates were oriented perpendicularly to the solar rays, then a cosine factor would be avoided increasing the efficiency in the reflection. 
Third, the metallic plates could be located on the roofs of the houses instead of on the ground. In extended cities this sounds a better option than in most European cities where the high density of housing and the contrasts in the heights of buildings would produce important shielding of the solar rays.
Fourth, we are here emphasizing the virtues of the metal plates, but obviously any other initiative such as painting roofs or roads with reflecting colours would be pushing in the same right direction. 
Fifth, and most important, the aim of increasing the Earth's albedo as proposed here is important for reducing the planetary temperature, but even if it were already operative this would not mean that present efforts to reduce energy use, whatever its origin, should be reduced. Stimulating the efficiency and safety of new renewable energy sources, the rational use of water, etc. should continue to be encouraged. Likewise, campaigns to raise public awareness on these issues should receive the highest attention in political institutions.

\begin{acknowledgements}
This work was supported in part by a Project of the Spanish Ministry of Education and Science.
\end{acknowledgements}

\end{document}